\documentclass[12pt]{iopart}

\usepackage{iopams}

\usepackage{stackengine}
\newcommand\textss[1]{\stackengine{.9ex}{}{\scriptsize#1}{O}{l}{F}{F}{L}}
\usepackage{graphicx}

\begin{document}

\title{LGAD Detector Concept for TOPSiDE Project}

\author{Kyung-Wook (Taylor) Shin\textss{1*}, Jose O. Repond\textss{2}, David Blyth\textss{2}, Jessica E. Metcalfe\textss{2}, Manoj Jadhav\textss{2}, Abraham Saiden\textss{1}, and Hartmut Sadrozinski\textss{1} }
\ead{tashin@ucsc.edu\textss{1*}}
\address{1156 High St, Santa Cruz, CA 95064, USA\textss{1}, 9700 S Cass Ave, Lemont, IL 60439, USA\textss{2}}
\vspace{10pt}
\begin{indented}
\item[]December 2021
\end{indented}

\begin{abstract}
We report a concept low gain avalanche diode (LGAD) detector to be integrated into TOPSiDE\cite{Repond:2018K9}, which is being developed for EIC (Electron-Ion Collider) project. The LGAD detector will be taking its place to resolve requirement of time-of-flight measurement for the TOPSiDE project. To achieve the required timing resolution of 10 $ps$, the LGAD will be completely monolithic structure, eliminating the parasitic elements from hybridization. The detector itself will be consist of the LGAD sensor, amplification and discrimination, and time-over-threshold logic components. At this stage, we are optimizing the LGAD sensor design and preparing a discrete component based readout circuitry. In this work, we will present simulation result of a conceptual LGAD silicon detector and constant-fraction discriminator (CFD) circuit implementation. The timing resolution of the conceptual detector was simulated with EDA tools such as Silvaco Atlas\textss{TM} and LTSpice\textss{TM} with conjunction of a Monte-Carlo simulator to resolve Landau distribution into the circuit readout simulation. The best timing resolution we achieved was 23 $ps$.
\end{abstract}

%
\vspace{2pc}

\section{Introduction}
Low gain avalanche diodes have been explored and adopted since 2012\cite{Hartmut:2012RD50}. Due to recent development of semiconductor industry, high quality p-type wafer supply became abundant so that a large scale silicon detector is not a financial and technological hurdle anymore. The semiconductor devices provide ample advantages compared to vacuum chamber based detectors due to their size (at least 3 orders of magnitude thinner than typical vacuum chamber.\cite{Wermes2016, spieler2005semiconductor}) Thus, greatly improves the detection speed as well as timing resolution which determines the accuracy of measurement of momentum, therefore, mass of sub particle. Although, a prolonged exposure to high-energy particles or photons may degrade the quality of silicon detector, limiting its life cycle, the ample supply of high quality wafers negate such drawback. The fundamental limitation of silicon wafers as a timing resolution detector has been theoretically suggested as 10 $ns$ range at best \cite{Riegler_2017} which is already enough for TOPSiDE application.

\begin{figure}[ht]
\centering
\includegraphics[width=0.75\textwidth]{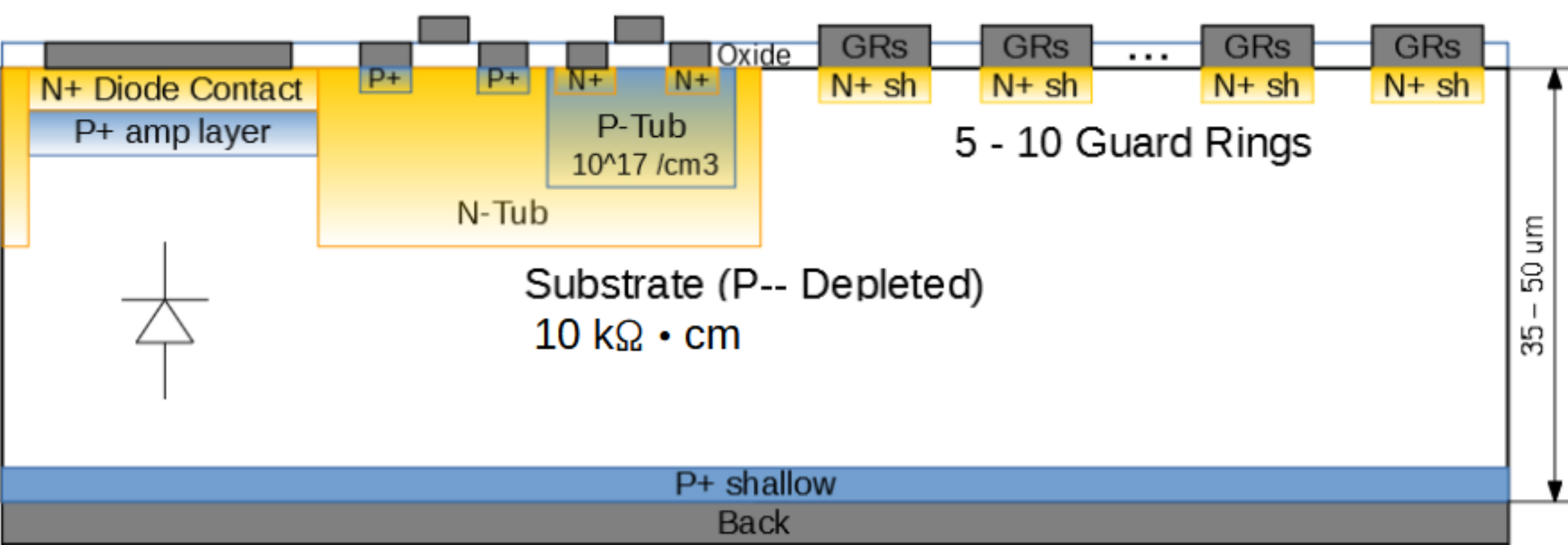}
\caption{Conceptual drawing of the monolithic LGAD detector.}
\label{fig:concept}
\end{figure}

We are approaching with a complete monolithic detector concept to take advantage of modern industrial semiconductor production capability. It is well known that many scientific ASIC projects worked with foundries to implement specific technical requirements. The LGAD sensor and its readout elements, including not only constant fraction discriminator but also time-over-threshold (TOT) implemented with time-to-digital converter(TDC)s, into a single pixel. The model pixel (for LGAD sensor simulation) of the conceptual device can be found in Figure \ref{fig:concept}. The single pixel model consist of LGAD sensor (leftmost side,) dummy MOSFET device models, and a set of guard rings (right side of the device model in the Fig \ref{fig:concept}.) 

In this work, TCAD simulation based on the model shown in Figure \ref{fig:concept} will be discussed to establish the conceptual device itself. The TCAD simulation includes not only electrical characterization of the LGAD sensor but also includes simulation results from simulated particle generation to establish the timing resolution. Accordingly, an ideal CFD circuit was implemented to evaluate the timing resolution to justify the timing resolution of 23 $ps$. The circuit simulation was performed with selected beam profile to mimic the Landau fluctuation in most of silicon based particle detectors.

\section{TCAD Simulation}
In this section, we will discuss on LGAD conceptual implementation. Since the development involves silicon device simulation, we have implemented Silvaco\textss{TM} Atlas as a main simulator. Due to the requirement of LGAD, the device itself was implemented as a regular P-i-N diode structure as depicted in Fig. \ref{fig:concept} with omission of CMOS device area (but leaving the N-Tub) to reduce simulation complexity. To implement the LGAD, which requires HV bias from the backside of the wafer, we have also implemented a set of guard rings. The LGAD in this conceptual implementation was 1 mm by width but 1 $\mu$m long since the simulator assumes every device to be such in terms of length.

The silicon substrate was assumed to be high resistivity P-- (10 $k\Omega \cdot cm$) depleted silicon wafer with 10\textss{12} $cm^{-3}$ of boron concentration. The n+ top contact layer has been implemented with a phosphorus Gaussian analytical profile of 50 $nm$ of standard deviation, peaking at the top surface with 10\textss{19} $cm^{-3}$ of concentration. Additionally, the top side contact has been extended with phosphorus peak with 10\textss{17} $cm^{-3}$ of peak concentration and 1 $\mu m$ of standard deviation to be paired with a deep boron implant, amplification layer. At the back side, the p+ layer was implemented with boron Gaussian analytical implementation, peaking 10\textss{17} $cm^{-3}$ of concentration with 1 $\mu$m of standard deviation from the back electrode. The JTE area was implemented with, again, phosphorus Gaussian analytical profile, peaking 10\textss{17} $cm^{-3}$ with 3 $\mu m$ of standard deviation from the top side. The full depletion was achieved at -87 V of back plate bias.

\subsection{Guard Ring Optimization}
Since a full depletion is required to collect electron-hole pair effectively, the back plate bias needs to be excessive, compared to CMOS operating range. Since this device was conceptual stage, we have taken account that the back plate bias can be as high as -200 V of bias to provide not only full depletion but also ample electric field to encourage 'soft' avalanche breakdown at the vicinity of the top electrode with an amplification layer which will be implemented as a boron analytical Gaussian profile.

\begin{figure}[ht]
\centering
\includegraphics[width=0.75\textwidth]{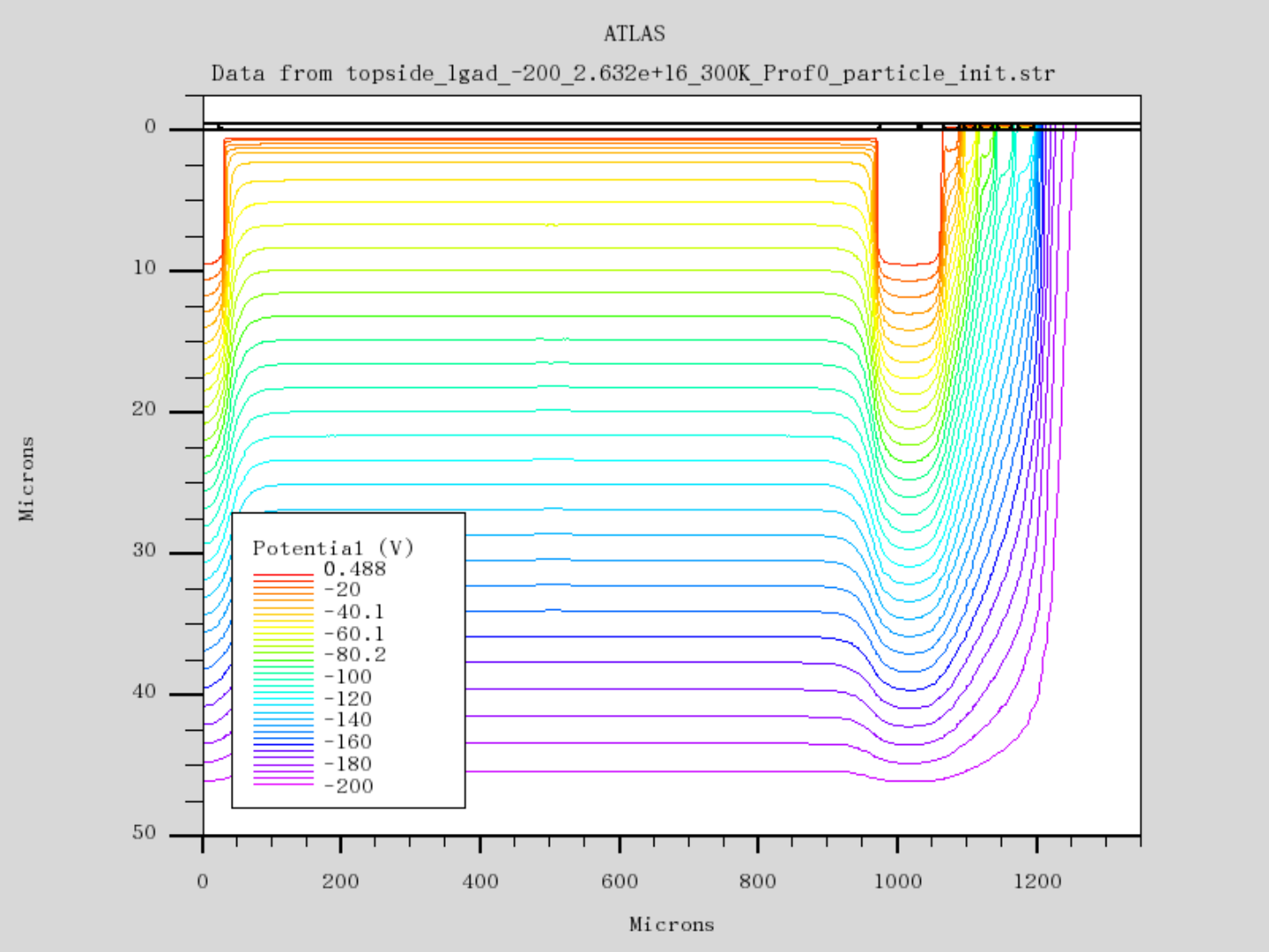}
\caption{Potential distribution under -200 V of back plate bias.}
\label{fig:Pot200VBias}
\end{figure}

However, biasing any isolated n+ contact electrode may cause early breakdown due to fringing field at the edge of implantation. Also, depletion region, which will be starting from the top n+ electrode as we increase the back plate in this case due to p-type substrate, needs to be suppressed to not to reach the die termination, we have implemented 5 of floating guard rings with 50 to 100 $\mu$m of increasing pitch from the LGAD vicinity, with 20 $\mu$m of spacing each. The guard ring contact implantation profiles are that of LGAD n+ contact. The potential distribution around the LGAD can be seen in \ref{fig:Pot200VBias} under -200 V of back plate bias. The preliminary guard ring implementation was deemed to be acceptable since suppressing the depletion was not reaching the die termination under -200 V effectively based on the potential distribution in the Fig. \ref{fig:Pot200VBias} where -200 V of potential curve does not expand into the right edge of the device.

\subsection{Amplification Layer}
\begin{figure}[ht]
\centering
\includegraphics[width=0.75\textwidth]{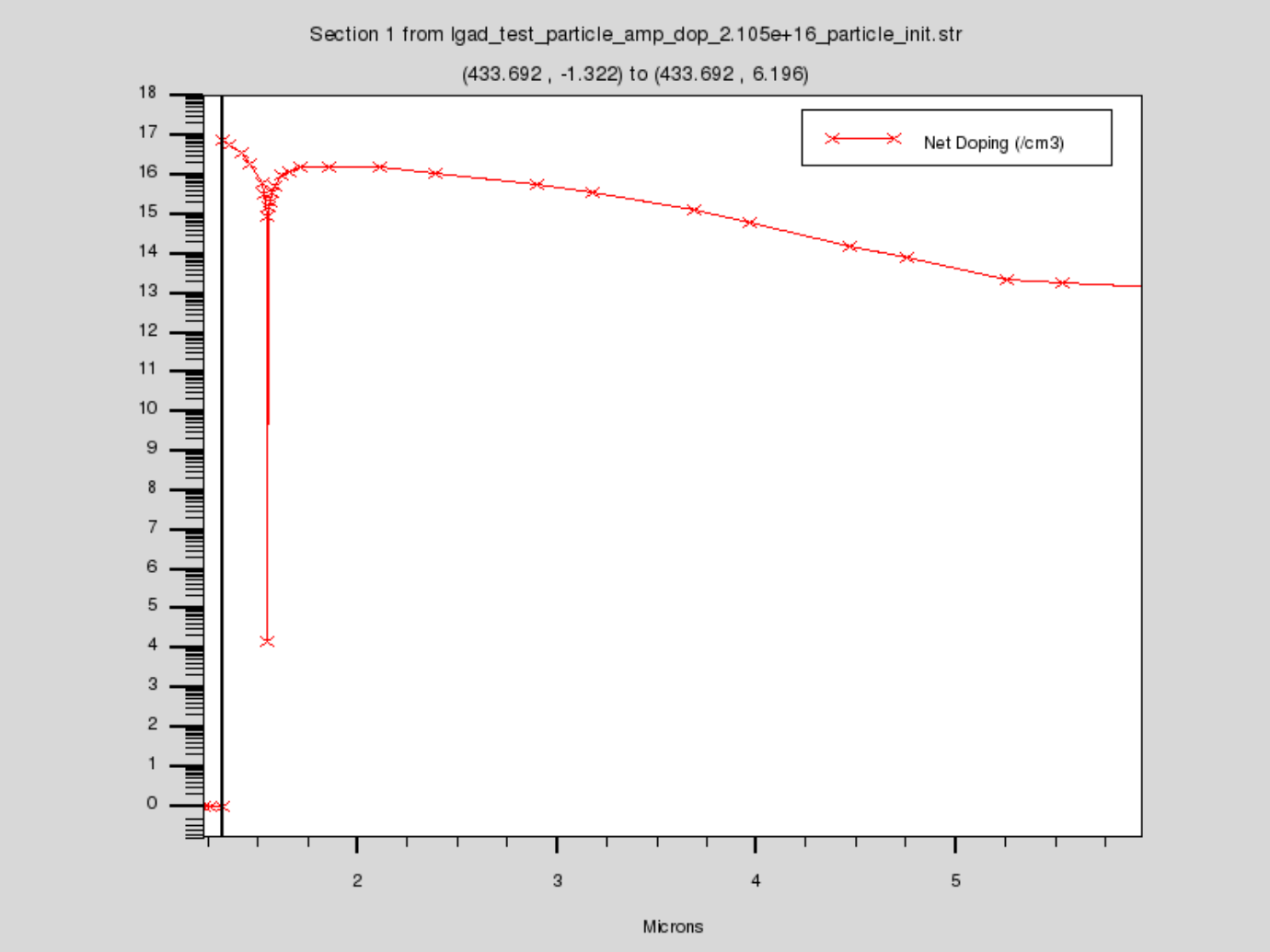}
\caption{Net doping concentration of top electrode tail and amplification layer at the top electrode vicinity. The contact vicinity concentration (10\textss{19} $cm^{-3}$) was ignored due to scaling.}
\label{fig:AmplDoping}
\end{figure}

The implementation of amplification layer involved iterative approach, utilizing the Design of Experiment (DOE) functionality. The amplification layer, again analytical Gaussian profile, was placed 1 $\mu$m deep from the top contact vicinity, with 3 $\mu$m of standard deviation. The DOE iteration was performed for the peak concentration, spanning from  
10\textss{16} $cm^{-3}$ to 10\textss{18} $cm^{-3}$. While running the iteration, the leakage current has also increased from 0.01 nA at the start of the loop to over 10 nA range at the end of iteration, which represents 10 $\mu$A of leakage current since we are conceptualizing 1 mm by 1 mm square pixel. Therefore, we decided the peak concentration of 2.6 $\times$ 10 \textss{16} $cm^{-3}$ as the amplification layer's peak concentration, resulting the net doping concentration at the top electrode vicinity shown in Fig. \ref{fig:AmplDoping}. We can find the phosphorus tail and the amplification layer here.

\begin{figure}[ht]
\centering
\includegraphics[width=0.75\textwidth]{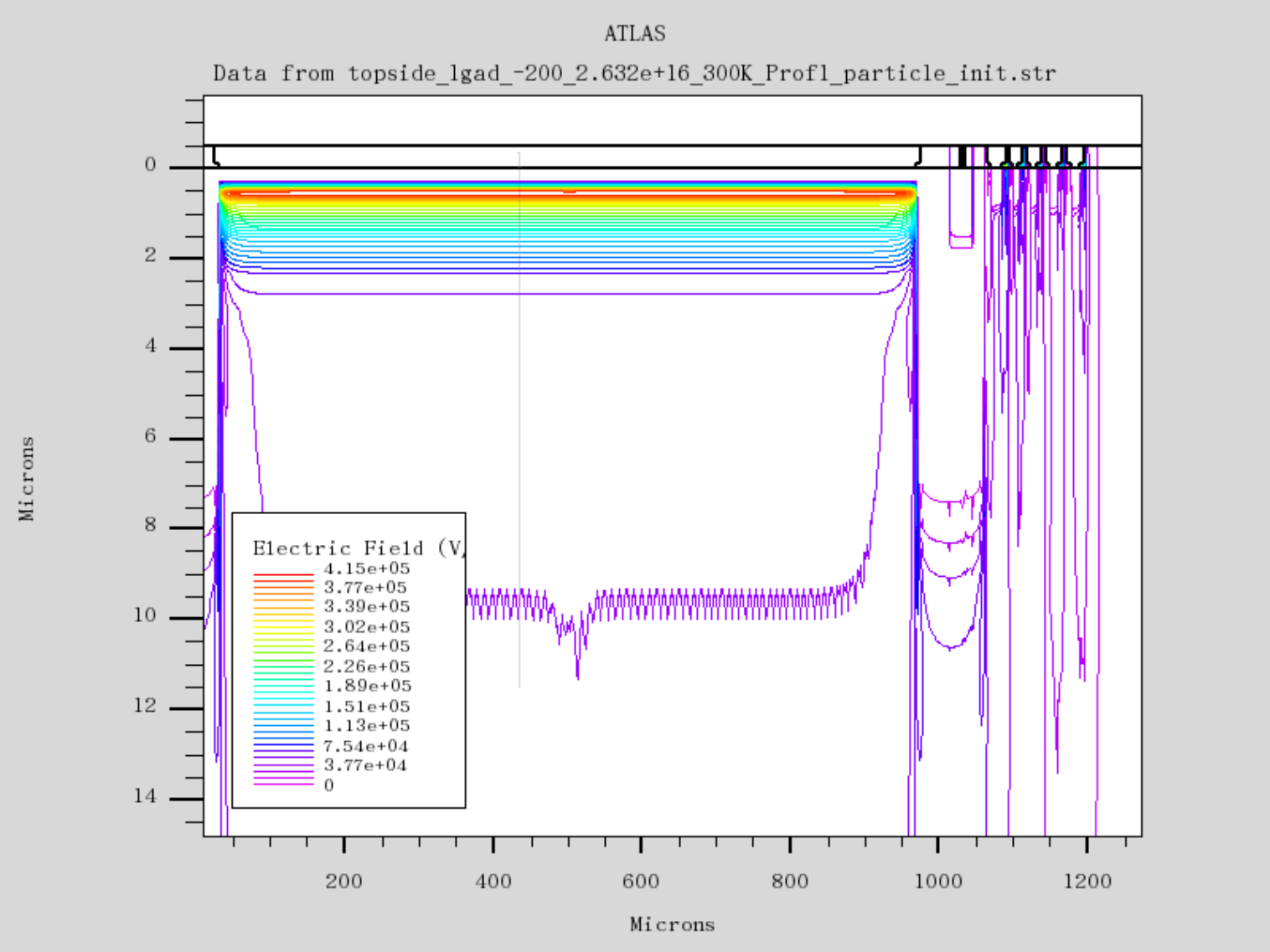}
\caption{Simulated electric field at the amplification layer vicinity.}
\label{fig:EFieldAt200VBias}
\end{figure}

To icite any breakdown in a TCAD simulation, we need to establish at least 10\textss{5} V/cm of electric field at the vicinity of any P-N junction. Although we have settled down at the lower end of DOE iteration, the electric field at the amplification layer vicinity was large enough to start any avalanche generation under -200 V of back plate bias as depicted in Fig. \ref{fig:EFieldAt200VBias}, resulting 4.15 $\times$ 10\textss{5} $V/cm$ as peak electric field, which was enough to start avalanche breakdown mechanism in the TCAD algorithm. 

The peak electric field can be adjusted by moving the amplification layer's peak location deeper or lowering the peak concentration of the amplification layer. However, we decided to settle down at 1 $\mu$m deep or shallower to stay within CMOS fabrication parameters. The little pinched out region at 500 $\mu$m (in the Fig. \ref{fig:EFieldAt200VBias}) is due to numerical mishap due to finer grid size at the location. The device mesh size at the site is an order of magnitude smaller than other areas to implement a particle beam which has virtually zero thickness. 

The avalanche breakdown (impact ionization) model we have selected was Okuto-Crowell since it was mostly adopted for various p-i-n diode simulations and also has temperature dependence factor since the sensor will be operating under 200-250 K in some situations. However, there are other impact ionization models such as Van Overstraeten de Mans or Massey models needs to be considered in future \cite{mandurrino2017tcad}. The simulated amplification can be found out in the later chapters (Fig. \ref{fig:tran_LGAD}) where transient response curves presented.



\section{Readout Circuit}

Even though the original conceptual device is supposed to mount electronic components, regarding trans-impedance amplifiers, discriminators, and time-of-arrival detection, into a single pixel, as a monolithic implementation, we were also experimenting the sensor elements separately. Thus, we have been also working on a discrete component based PCB readout system in parallel to sensor design optimization in TCAD simulation. Hence, the readout simulation of the LGAD detector system is based on the discrete component readout board design. The section below will be describing the ideal readout circuitry which will be implemented as a discrete component readout system, later.

\subsection{Constant fraction discriminator}

\begin{figure}[ht]
\centering
\includegraphics[width=0.85\textwidth]{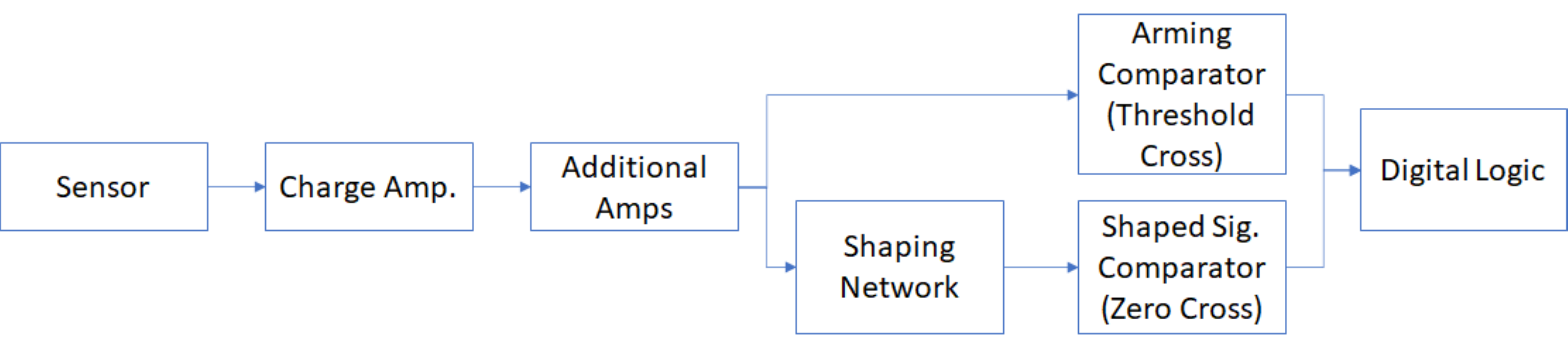}
\caption{LGAD detector readout chain.}
\label{fig:readout_chain}
\end{figure}

The idea of constant fraction discriminator is widely employed to ensure timing resolution as best as possible since 1960s \cite{osti_4520252, KIELEK1996392, AlShamsi:2651090}. Fig. \ref{fig:readout_chain} shows a typical readout chain of timing resolution detectors. Due to recent developments in semiconductor industry, we can even remove additional amplifier stages after charge amplifier (or transimpedance amplifier) stage if we are implementing this circuitry into the silicon wafer. However, the board we have designed was based on display driver amplifiers which resulted a couple of additional stages of amplifiers.

\subsubsection{Preamplifier}

The preamplifier part of readout was implemented with three stage amplifiers, including a trans-impedance amplifier stage with trans-gain of 40. The amplifier we have chosen to drive all the stages is a display driver component, AD8009 from Analogue Devices\textss{TM}. The feedback capacitor was chose as 2 pF to be paired with 1 k$\Omega$ of feedback resistor, ensuring around 10 MHz of operational frequency before the 3 dB roll-off. The sensor is AC coupled from relatively large, 0.1 $\mu$F capacitor, however, DC coupling can also be set up without the AC coupling capacitor. Since our sensor has 2.02 pF of internal capacitance, the sensor input time delay can be as low as 50.5 ps which is small enough, compared to sensor readout time, around 1.5 ns, depicted in Fig. \ref{fig:tran_LGAD}.

\begin{figure}[ht]
\centering
\includegraphics[width=0.8\textwidth]{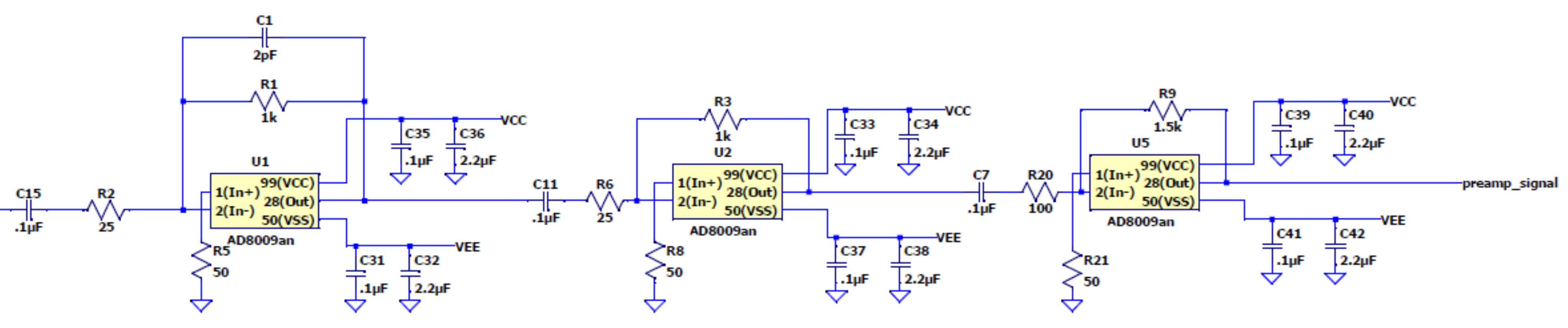}
\caption{Preamplifier stage.}
\label{fig:preamp}
\end{figure}

Additional amplifiers are also AC coupled to provide additional V-to-V amplification until reaching the shaper and discriminators for time of arrival validation. The entire trans-impedance amplification ratio is 24000. The entire preamplifier network can be found in Fig. \ref{fig:preamp}.

\subsubsection{Shaper}

\begin{figure}[ht]
\centering
\includegraphics[width=0.55\textwidth]{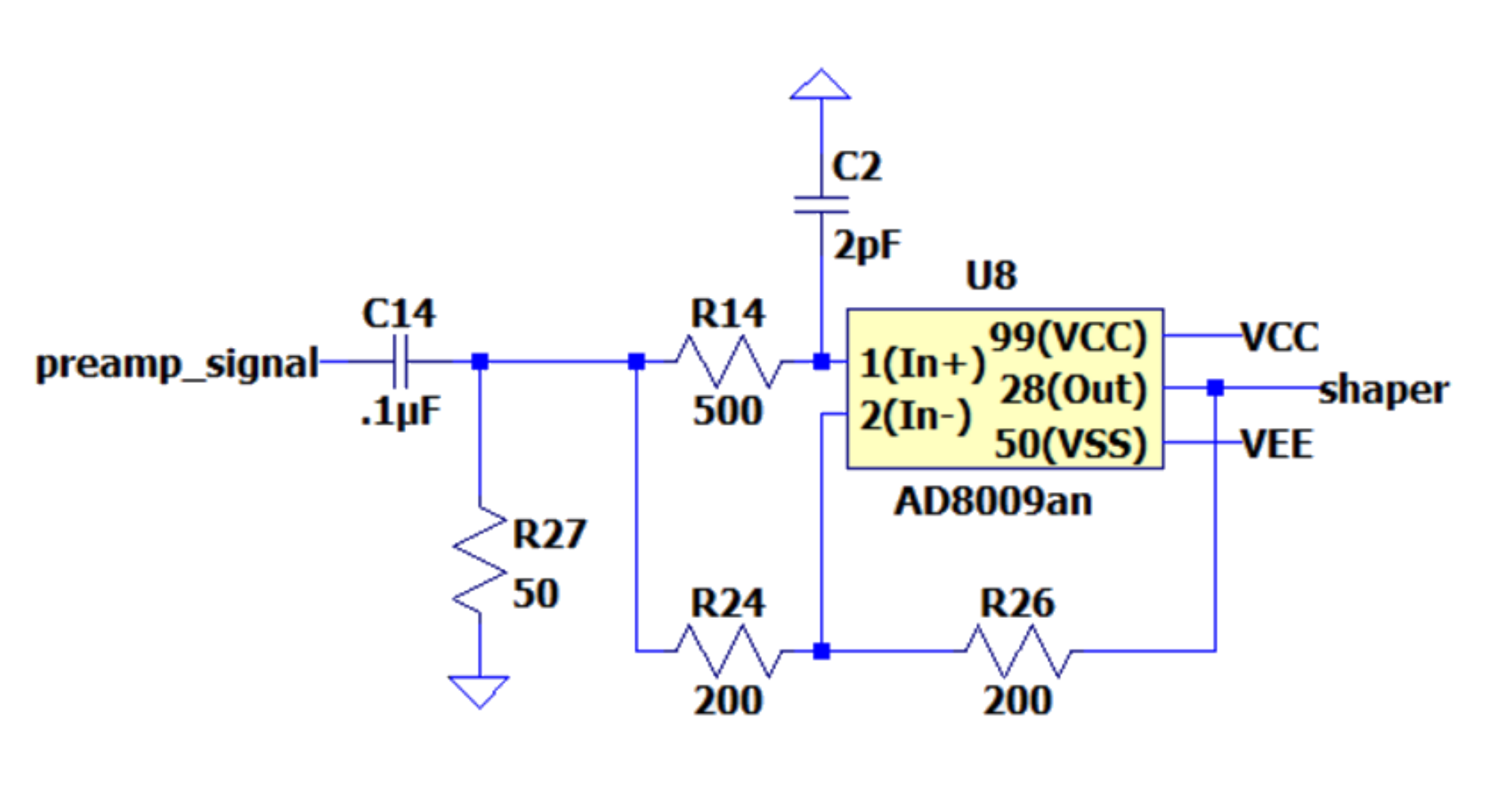}
\caption{Shaping circuitry.}
\label{fig:shaper}
\end{figure}

The shaper network, depicted in Fig. \ref{fig:shaper}, is designed rather than implementing a couple of delay and fully functional circuitry but using a phase amplifier, using the same op-amp component as the preamplifier stage. The phase of the shaping network will be programmable with given input resistors which will be selected through a high-speed, low impedance multiplexers (ADG1604.) The shaper parameter was tuned to facilitate the timing resolution simulation, in terms of tuning the constant fraction discriminator, the fraction of 44 \% and 4 ns of delay. The fraction and delay parameters will be adjusted by manipulating the resistors: R14 and R24 in production circuitry using multiplexers. The 50 $\Omega$ shunt resistance (R27) is installed there to drain remaining charges from the AC coupling capacitor C14.

\subsubsection{Time of Arrival}

\begin{figure}[ht]
\centering
\includegraphics[width=0.8\textwidth]{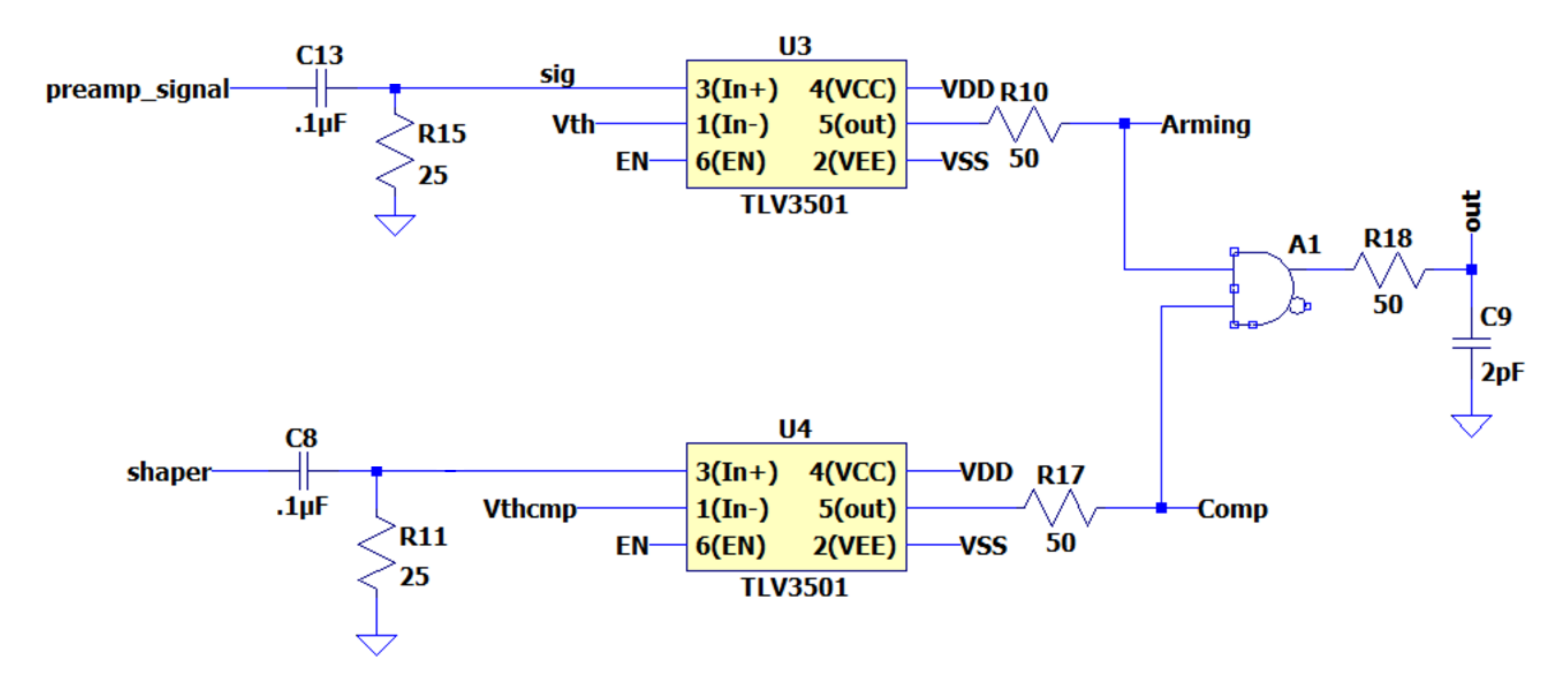}
\caption{Time of arrival detection.}
\label{fig:TOA}
\end{figure}

Time-of-Arrival circuitry is composed of two comparators with digital output (TLV3501,) each serves as its own purpose as depicted in Fig. \ref{fig:TOA}. The arming comparator provides rough information on the arrival of particle while the shaped signal actually provides the arrival time to be collected to figure out the timing resolution. Since we are just using the shaped signal to determine the timing resolution, both signals are tied to an ideal AND gate. The arming signal comparator part is implemented to provide systematic figure since we will be needing the arming signal to operate TDC. The timing signal is collected at an arbitrary RC network, implemented with R18 and C9. The threshold voltages for comparators are 0.1 V for arming comparator and 16 $mV$ for shaped signal (which implemented to be close to zero cross,) respectively.

\subsection{Circuit Simulation Scheme}

\begin{figure}[ht]
\centering
\includegraphics[width=0.65\textwidth]{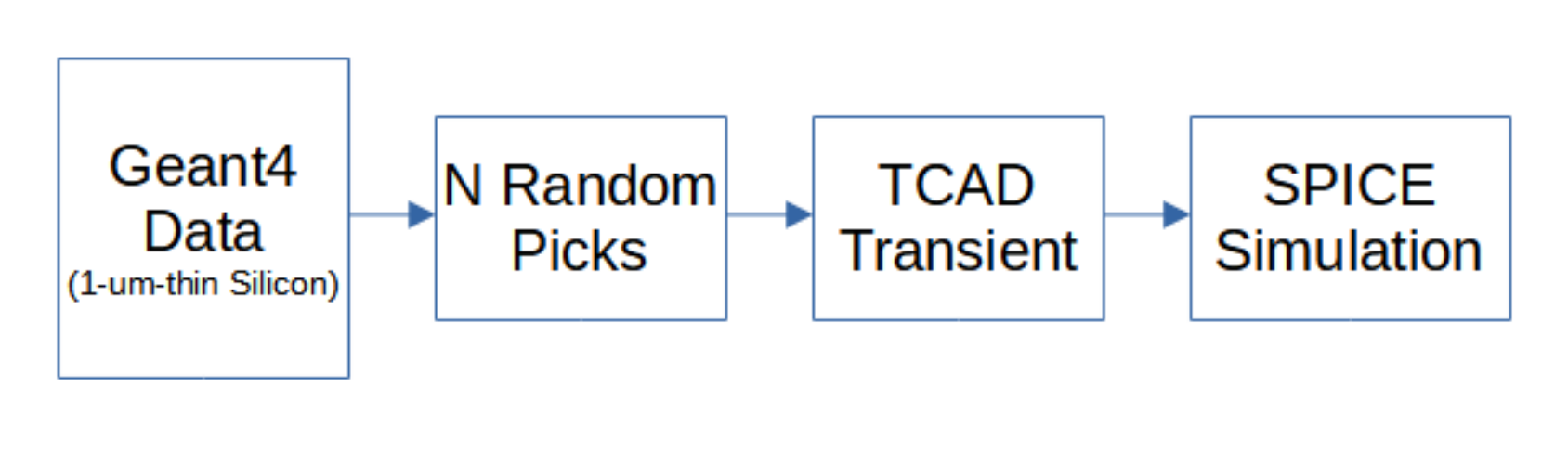}
\caption{Timing resolution simulation scheme.}
\label{fig:simul_scheme}
\end{figure}

To evaluate timing resolution of the LGAD we have simulated with the TCAD simulator, we have developed a simple implementation of constant fraction discriminator system as described in previous chapter. The timing resolution can only be simulated relying on statistical approach to involve the Landau distribution of the silicon detector itself. To imply this phenomena, we have implemented a little chain of simulation as depicted in Fig. \ref{fig:simul_scheme}.

\begin{figure}[ht]
\centering
\includegraphics[width=0.65\textwidth]{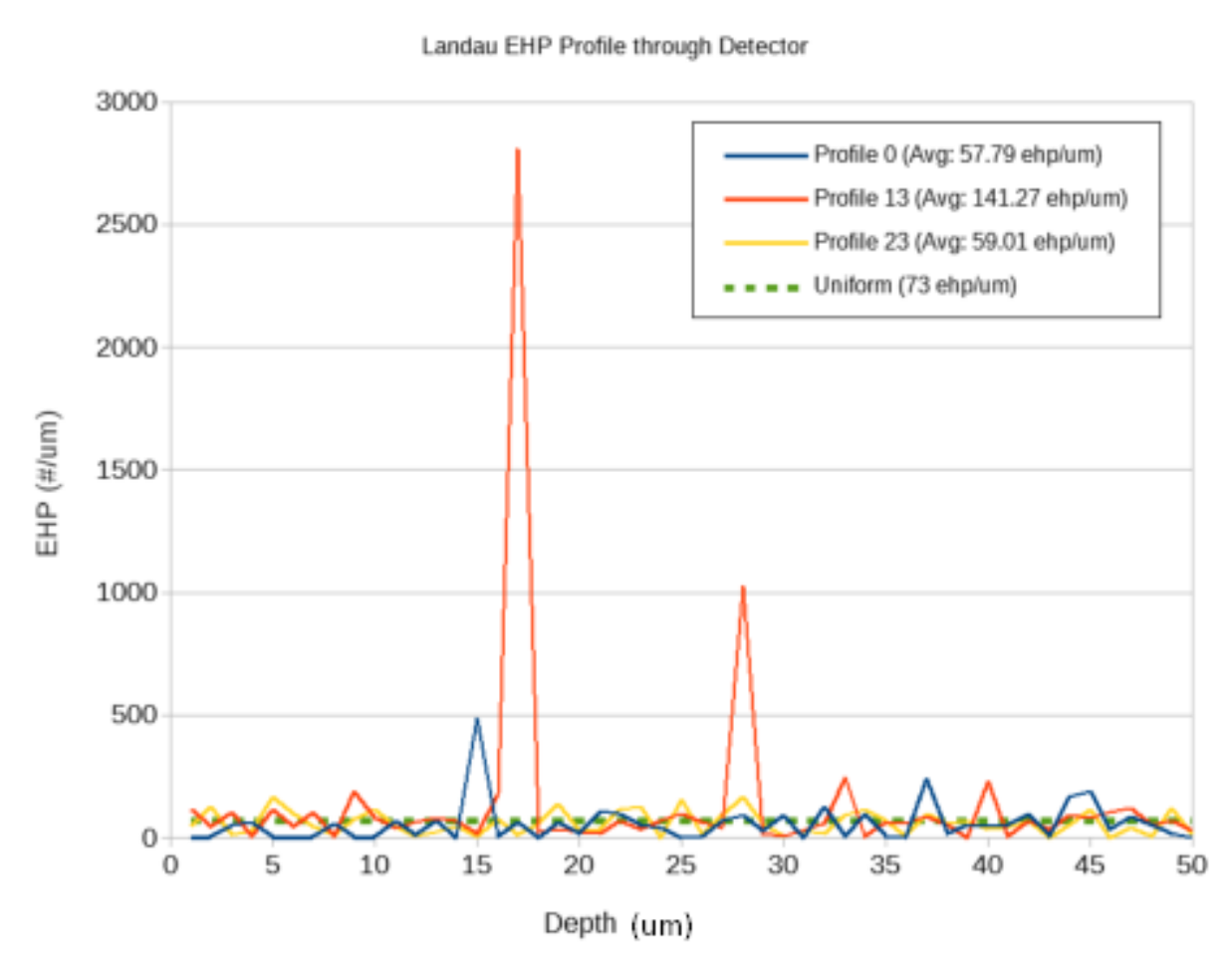}
\caption{A few Landau profiles across the 50-$\mu$m-thick LGAD.}
\label{fig:landau_profiles}
\end{figure}

\begin{figure}[ht]
\centering
\includegraphics[width=0.65\textwidth]{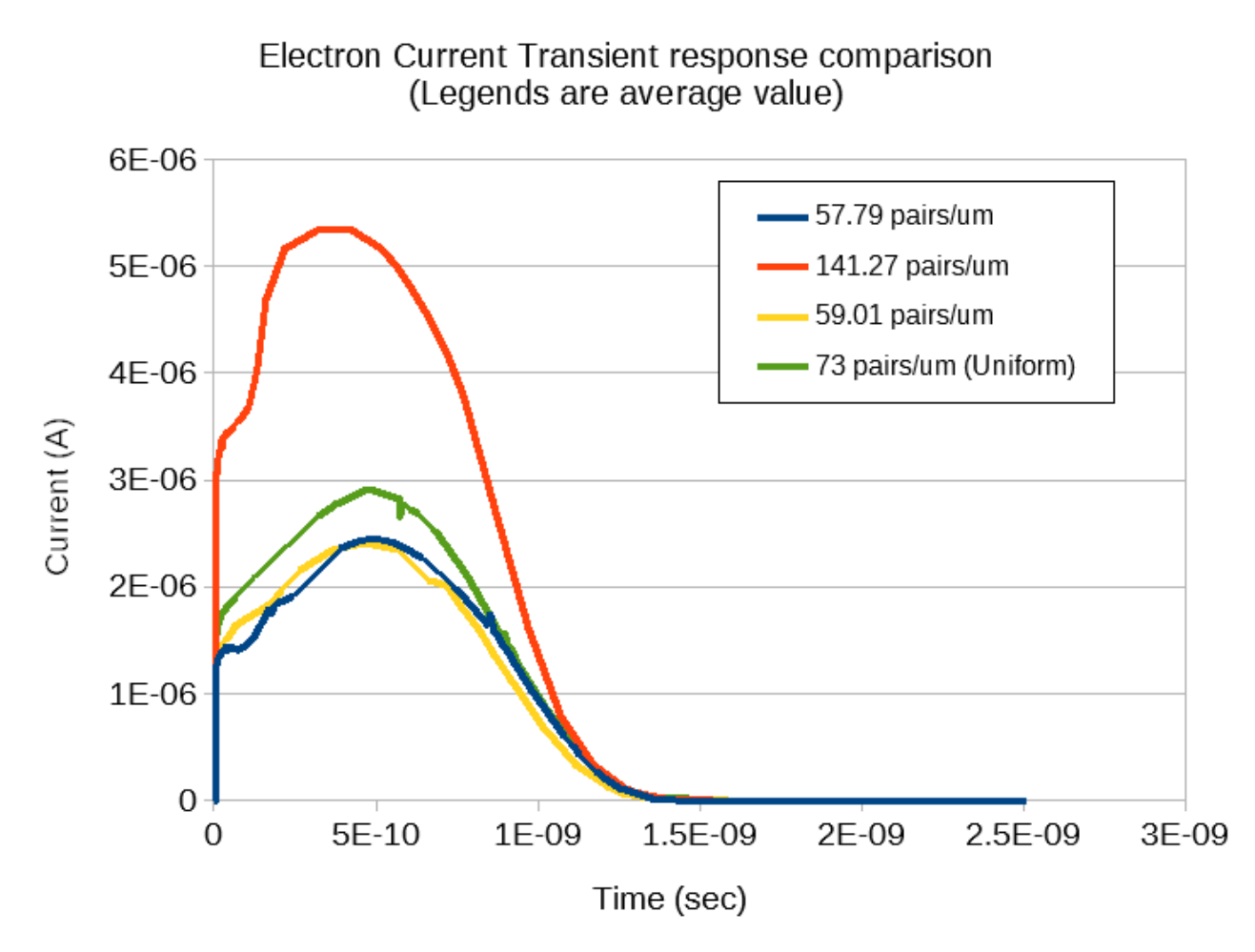}
\caption{Transient response from the particle generation situations depicted in Fig. \ref{fig:landau_profiles}.}
\label{fig:tran_LGAD}
\end{figure}

At the first stage, due to statistical nature of electron-hole pair generation in silicon, we have assumed a 2 $GeV$ muon penetration on a 1-$\mu$m-thick silicon chunk in Geant4 simulator to gather more than 10,000 electron-hole pair generation cases. Since our detector is only 50-$\mu$m-thick, all that we had to pick out of 10,000 data was 50 samples. The sample selection was indeed relied on random number generator. Fig. \ref{fig:landau_profiles} shows a few exempt cases of the sample picking, compared to 73 MPV/um case that we have characterized the LGAD device model against. The LGAD device response, prior to be fed into the constant-fraction discriminator circuitry, can be found in Fig. \ref{fig:tran_LGAD}. 

We have made 200 cases of 50 1-$\mu$m-thick silicon electron-hole pair generation random picks to generate a statistical data of a simulated LGAD readout. We have ran 200 TCAD simulations for each cases of random picks of electron-hole pair generation, then the 200 device current has been fed into the CFD circuitry to be simulated with a SPICE simulator. We have used LTSpice\textss{TM} to simulate the circuit response of each random pick case and picked up the excitation time at the output port of the comparator network where the output port of the ideal AND gate provides logic signal, switching from 0 to 1, when the shaped signal crosses zero threshold voltage which is tuned to 16 mV. The statistical result can be found in Fig. \ref{fig:timing_res}.

\section{Results and Discussion}

\begin{figure}[ht]
\centering
\includegraphics[width=0.65\textwidth]{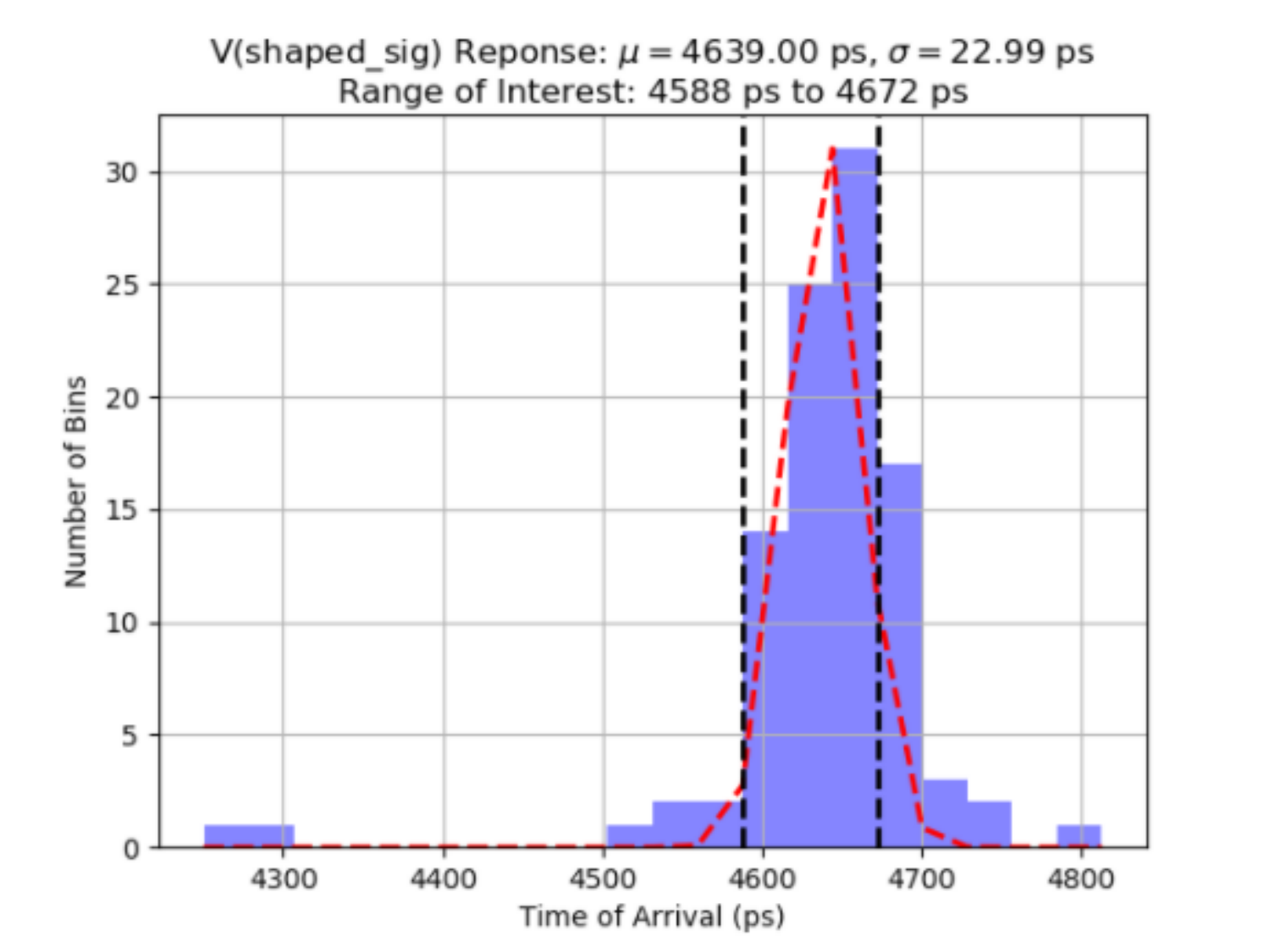}
\caption{Simulated timing resolution of the LGAD detector concept.}
\label{fig:timing_res}
\end{figure}

As depicted in Fig. \ref{fig:timing_res}, we can suspect the timing resolution using discrete component CFD would be as much as 22.99 ps. Indeed it is not an ideal data since the readout circuitry itself was not prepared to be a high performance integrated circuitry but rather put together components. Also, this data set is missing interference from circuit noise components. However, we suspect that the noise component will not be playing critical role due to amplification ratio of 24,000 at the pre-amplifier stage and the leakage current of the LGAD device stayed at 10 $pA$ range while the photo-excited electron-hole pair current from the LGAD was at least 5 orders of magnitude higher than the leakage current. Since the timing resolution is mainly determined by jitter, which is inversely proportional to signal-to-noise ratio (SNR) of a LGAD detector system, we already have a massive SNR due to three stage pre-amplifier stage which may overwhelm the other noise sources.

The limitation of BJT based discrete component is currently preventing higher timing resolution to be reached. Thus, resulting early 20 $ps$ range at the best. In the monolithic implementation, the amplification and shaping circuitry will be replaced with integrated circuitry which would have much smaller effect from parasitic components. Therefore, we can expect better timing resolution ($<$ 20 $ps$) than as of now. However, since we also need to integrate noise components of time-of-arrival circuitry, especially time-to-digital converter part, it may end up more than 23 $ps$ range once the noise factors are included into the simulation model. Especially, a typical TDC is composed with hundreds or thousands of flip flop elements, the number of elements and resolution depends heavily on a fabrication process, is expected to add up more uncertainty to actual readout data. At this time, the TDC is planned to be implemented with a Xilinx or Intel FPGA unit which already shows promising timing resolution in other works \cite{s17040865, sven20201, kuang20185}. 

\section{Conclusion}

In this work, we have conceptualized a monolithic LGAD detector for particle experiment. The detector part was implemented with a LGAD based on industry standard material, silicon, and evaluated its timing resolution with an semi-ideal CFD based readout circuitry. The timing resolution from the circuitry was around 23 ps which provides a starting point for further improvement with higher transimpedance gain and bandwidth benefited from monolithic implementation. Thus, further work will be needed to implement the integrated analog circuitry based on a MOS or BJT (even including SiGe) processes, comparable to most vendors provide. 

\section{References}
\bibliographystyle{unsrt.bst}
\bibliography{TOPSiDE_Paper}

\end{document}